\documentclass[]{spie}  %>>> use for US letter paper
\usepackage[square,sort,comma,numbers]{natbib}
\usepackage{amsmath,amsfonts,amssymb}
\usepackage{graphicx}
\usepackage[colorlinks=true, allcolors=blue]{hyperref}
\usepackage{siunitx}
\usepackage{sidecap}
\sidecaptionvpos{figure}{c}
\usepackage{wrapfig}
\usepackage{aas_macros}
\usepackage{subfig}

\providecommand{\adsurl}[1]{\href{#1}{ADS}}

\title{\href{https://spie.org/AS24/conferencedetails/astronomical-interferometry}{The Big Fringe Telescope}}

\author[a]{Gerard T. van Belle}
\author[b]{Anders M. Jorgensen}

\affil[a]{Lowell Observatory, 1400 West Mars Hill Rd., Flagstaff, AZ 86001, USA}
\affil[b]{Electrical Enginering Department, New Mexico Institute of Mining And Technology, 801 Leroy Pl., Socorro, NM 87801, USA}

\authorinfo{Further author information: (Send correspondence to G.v.B.)\\G.v.B.: E-mail: gerard@lowell.edu, Telephone: +1 928 233 3207}

\begin{document} 
\maketitle

\begin{abstract} %200-300 words
The Big Fringe Telescope (BFT) is a facility concept under development for a next-generation, kilometer-scale optical interferometer.  Observations over the past two decades from routinely operational facilities such as CHARA and VLTI have produced groundbreaking scientific results, reflecting the mature state of the techniques in optical interferometry.  However, routine imaging of bright main sequence stars remains a surprisingly unexplored scientific realm.  Additionally, the three-plus decade old technology infrastructure of these facilities leads to high operations \& maintenance costs, and limits performance.  We are developing the BFT, based upon robust, modern, commercially-available, automated technologies with low capital construction and O\&M costs, in support of kilometer-scale optical interferometers that will open the door to regular `snapshot' imaging of main sequence stars.  Focusing on extreme angular resolution for bright objects leads to substantial reductions in expected costs through use of COTS elements and simplified infrastructure.

\end{abstract}

\keywords{
optical interferometry,
main sequence star surface imaging,
starspots,
resolved exoplanet transit imaging,
solar analogs,
observing facilities,
microarcsecond astronomical imaging
}

\section{Introduction}\label{sec-introduction}

Astronomical interferometry in the optical has enjoyed a decade-plus of mature science operations.  %[don't bury the lede, so:]
Advancing further, from millarcsecond to microarcsecond scale imaging, is a readily achievable goal if a modern generation of tools are developed for this science using robust, recent technology.  Surprisingly, the near-immediate achievability, affordability, and merits of microarcsecond optical imaging of bright targets is low-hanging scientific fruit that has been overlooked until now.

Science results from the GSU CHARA Array\footnote{Georgia State University Center for High Angular Resolution Astrophysics Array} \cite{McAlister2005ApJ...628..439M,tenbrummelaar2005ApJ...628..453T} have included not just direct imaging but actual movies of the disk `finger' transiting $\epsilon$ Aurigae \cite{Kloppenborg2010Natur.464..870K}, the expanding fireball of Nova Delphini 2013 \cite{Schaefer2014Natur.515..234S}, and the surface of $\zeta$ Andromedae as it rotates \cite{Roettenbacher2016Natur.533..217R}.    The the ESO VLTI\footnote{European Southern Observatory Very Large Telescope Interferometer} \cite{Haguenauer2012SPIE.8445E..0DH} has imaged the surface of $\pi^1$ Gruis \cite{Paladini2018Natur.553..310P}, probed relativistic, non-Keplerian orbits of objects near our galactic center \cite{GRAVITY2020A&A...636L...5G}, as well as revealing the dusty veil that dimmed Betelgeuse \cite{Montarges2021Natur.594..365M}.   CHARA has also imaged the non-spherical surface of Altair \cite{Monnier2007Sci...317..342M}, and with VLTI, young stellar disks \cite{Kraus2020Sci...369.1233K}.  All of the studies cited here have resulted in marquee publications in either \textit{Nature} or \textit{Science}, or played a role in a Nobel Prize: the results noted here are just skimming the cream of a much larger body of work over the past decade that has made significant, unique contributions to astrophysics.

Overall, the technology infrastructure at CHARA and VLTI have realized the promise of extreme angular resolution from optical interferometry at the hundreds of meters scale.  However, those foundations are surprisingly aged.  For example, the delay lines of CHARA - the very heart of such an array - are literally from blueprints that are more than three decades old \cite{Colavita1992ESOC...39.1143C}; the VLTI delay lines are only slightly younger \cite{Messenger1998Msngr..91...25.}.   These tools reflect the state-of-the-art for their era, and the scope of the facilities, and have realized impressive scientific results.  However, they represent capabilities that were barely achievable with the technology at hand at that time, and their significant operations and maintenance overheads reflect those circumstances.

Carrying on from these grand achievements, we propose a new set of tools that will achieve a three-pronged set of goals.  First, we will take advantage of 3 decades of advancing technology. Second, our architectural design will aim for kilometric-scale optical interferometry, importantly enabling imaging science at the microarcsecond scale, including routine main sequence star surface imaging.  Finally, these tools will be designed within a framework of substantially reduced capital construction and operational / maintenance costs - again, leveraging improved technology as well as now-abundant operational experience from the existing facilities. %While these developments are independent of a specific observatory architecture we have in mind the Big Fringe Telescope (BFT) concept which is shown in Figure~\ref{figure_bft}, and it is described in more detail in \S \ref{section_bft}.

\begin{figure}
  \centering
  \subfloat[a][A `butterfly diagram' for the Sun, representative of the sunspot area coverage as well as the sunspot migration over time in latitude.  The BFT will be able to make such diagrams for exoplanet hosts, solar analogs, as well as a large number ($>800$) of general targets.]{\includegraphics[width=0.95\linewidth]{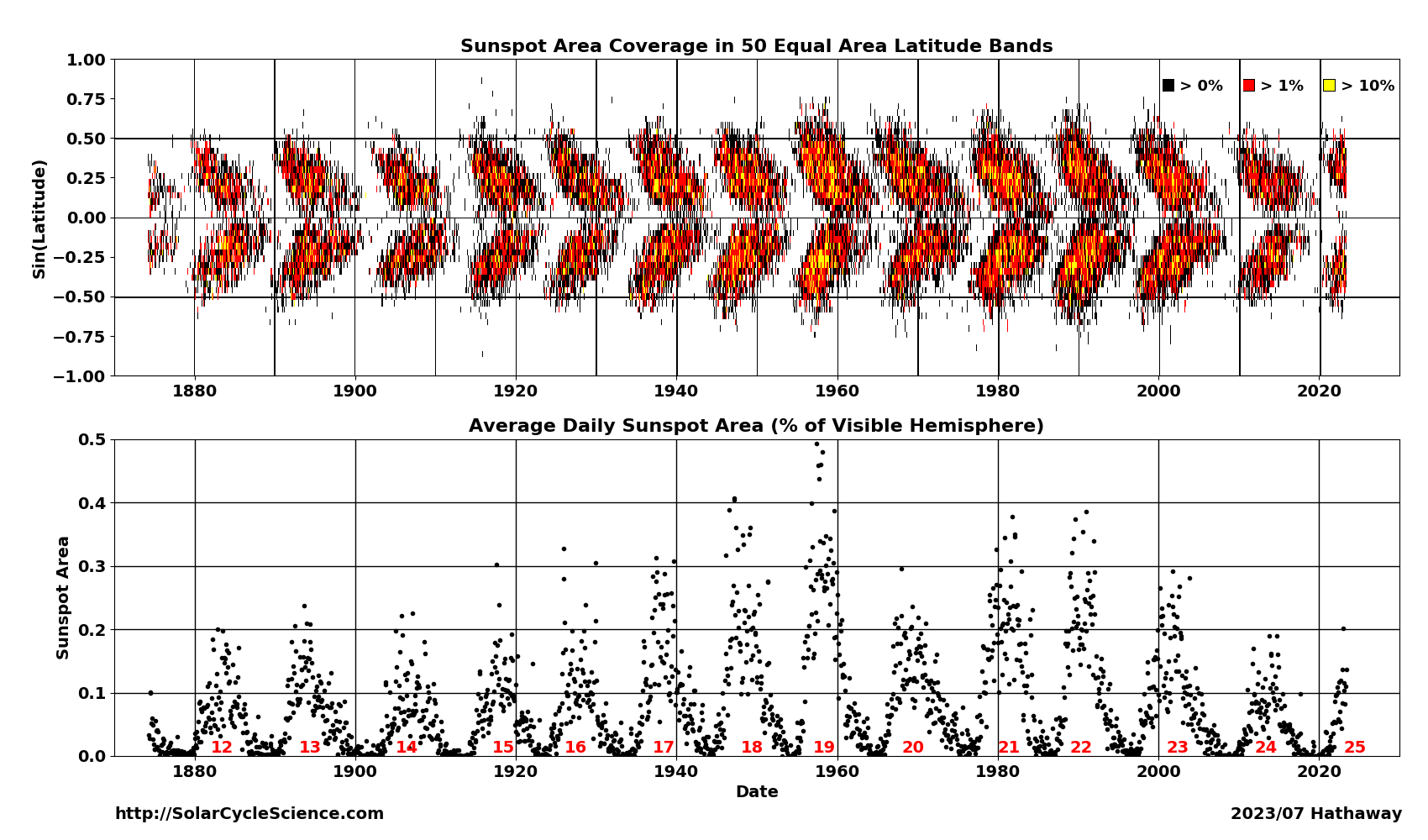} \label{fig-butterfly}} \\
  \subfloat[b][Venus transiting the Sun.  The BFT will be able to make resolved movies of such events for dozens of exoplanet hosts.]{\includegraphics[width=0.95\linewidth]{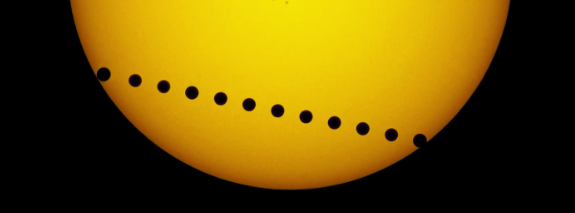} \label{fig-transit}}
  \vspace{5pt}
  \caption{Two of the key science applications of the BFT.} \label{figure-BFT_science_cases}
\end{figure}

\section{Science Case}\label{section_bft-science_case}

Four marquee science cases drive the architecture of the BFT, all based on the surprising fact that bright main-sequence star imaging remains a largely unrealized area of discovery space in astronomy.  The `sweet spot' of capabilities of BFT are as follows:
\begin{itemize}
    \item Optimal imaging of targets with angular sizes 0.40 to 0.60 mas (with imaging possible outside this range)
    \item Visible light sensitivity ($I$-band, 790nm) of 7.6 - 8.0 mag (requirement / goal)
    \item Near-infrared sensitivity ($H$-band, 1550nm) of 5.6 - 6.0 mag (requirement / goal)
    \item Resolving power of 0.20 - 0.90 $\mu$as (Michelson vs. Airy criteria - eg. 0.25 $\lambda$ / B vs. 1.22 $\lambda$ / B, `modeling' versus `true imaging'), which will enable surface characterization at the $10 \times 10$ to $30 \times 30$ pixel level for 0.40 to 0.60 mas targets, which is expected to be sufficient for detecting and monitoring magneto-hydrodynamically controlled processes manifesting on the stellar surface \cite{Carpenter2009ASPC..412...91C}.
\end{itemize}
In achieving these resolution and sensitivity goals, there are four primary science cases for the BFT.  Notably, these science cases known target lists in the range of $\delta\ = \{ -10^o,90^o\}$:

(1) {\bf Exoplanet hosts} being surveyed at the Lowell Discovery Telescope by the EXPRES precision-RV spectrograph \cite[e.g. Table 1 of][]{Brewer2020AJ....160...67B} are of particular interest, given the interest in correlating RV jitter with surface activity \cite{Roettenbacher2022AJ....163...19R}; 31 EXPRES targets are accessible in our BFT point design.  Many of these targets are also being monitored by other extreme precision RV spectrographs as well.

(2) Extending the stellar surface imaging to {\bf solar analogs}, a BFT telescope would be able to directly monitor the most Sun-like stars for their monthly and yearly surface morphology evolution (Figure~\ref{fig-butterfly}); there are at least 35 solar analogs \cite{Radick2018ApJ...855...75R} that are sufficiently large, bright, and in the northern hemisphere.  

(3) {\bf Resolved binaries} that can have both their orbits resolved,  as well as their individual component disk sizes resolved, will be able to establish empirically the mass-radius relationship; the SB9 catalog \cite{Pourbaix2004A&A...424..727P} has 77 known targets.  

(4) Finally, and most impressively, the BFT will be able to make real-time movies of {\bf resolved exoplanet transits} (Figure~\ref{fig-transit}), spatially resolving both the host star disk and the exoplanet itself.  The TFOP database \cite{Akeson2019AAS...23314009A} currently has 29 targets for which the star and planet are sufficiently resolvable and bright for our BFT point design.  

In addition to these four marquee cases spanning over 170 targets, there are roughly 800 general targets in the Bright Star Catalog \cite{Hoffleit1991bsc..book.....H}, spanning all spectral types except M, and all luminosity classes, which should be surface-resolvable (Tables~\ref{tab-general_targets-A},~\ref{tab-general_targets-B}) in the 0.30 - 0.70 mas angular size regime.  Bright stars figure prominently in the science case for the Habitable Worlds Observatory -- e.g. all of the 164 targets suggested by \cite{Harada2024ApJS..272...30H} for HWO are within BFT's sensitivity limits with $V<8$.

\begin{figure}
  \centering
  \subfloat[a][CHARA-MIRC imaging of the dusty disk of $\epsilon$ Aurigae eclipsing its primary star \cite{Kloppenborg2010Natur.464..870K}]{\includegraphics[width=0.95\linewidth]{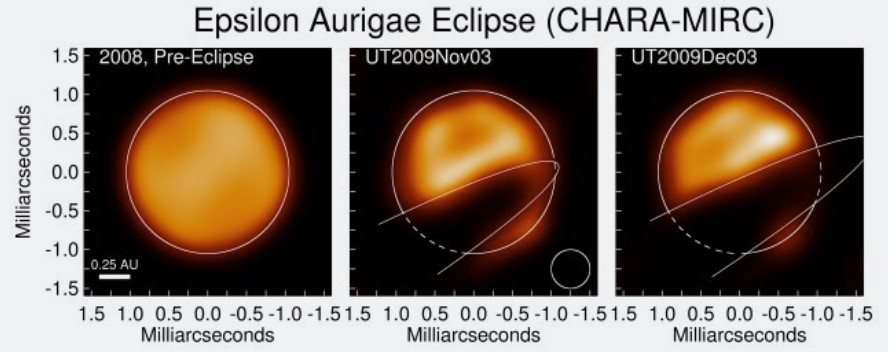} \label{fig-epsAur}} \\
  \subfloat[b][VLTI-GRAVITY imaging of the relativistic, non-Keplerian motion of star S29 approaching the supermassive black hole Sgr A* at the center of the Milky Way at a distance of only 90 AU \cite{GRAVITY2020A&A...636L...5G}.]{\includegraphics[width=0.95\linewidth]{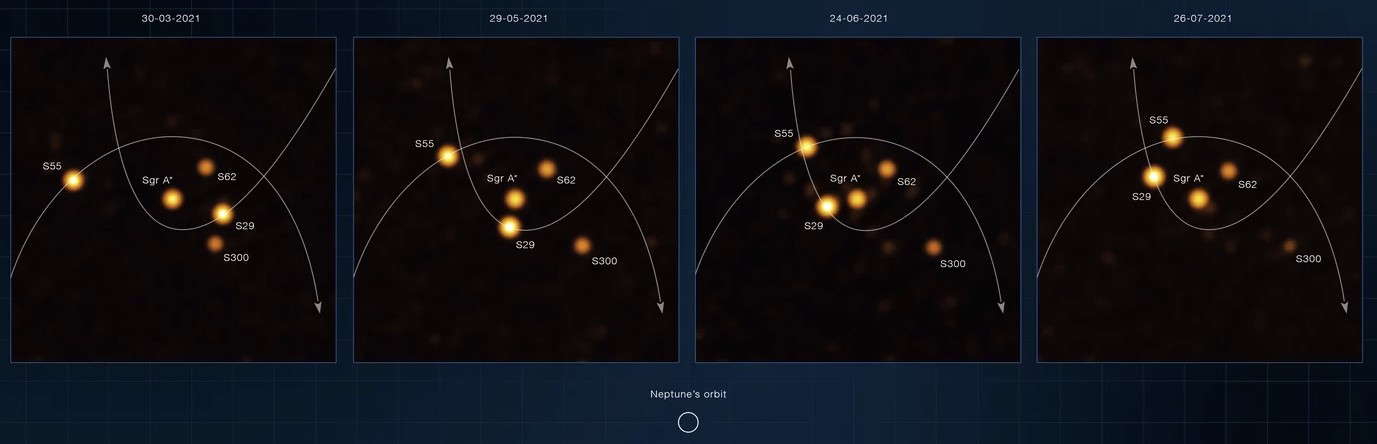} \label{fig-VLTI_SgrAstar}}
  \vspace{5pt}
  \caption{Modern astronomical optical interferometry imaging, reflecting the mature state of the technology.} \label{figure-interferometry_results}
\end{figure}

\begin{table}[!htb]
    \begin{minipage}{.5\linewidth}
      \centering

        \begin{tabular}{cc}
        \textbf{Sptype} & \textbf{N} \\
        \hline
        O               & 7          \\
        B               & 123        \\
        A               & 304        \\
        F               & 294        \\
        G               & 61        \\
        K               & 12         \\
        M               & 0     \\
        \hline
        \end{tabular}%
      \vspace{5pt}\caption{BFT targets by spectral type}\label{tab-general_targets-A}

    \end{minipage}%
    \begin{minipage}{.5\linewidth}
      \centering

\begin{tabular}{ccccc}
\textbf{LC} & \textbf{N} & \textbf{R}     & \textbf{d (pc)} & \textbf{$\theta$ (mas)} \\
\hline
1.0         & 23         & 37.4           & 905          & 0.511             \\
1.5         & 2          & 28.2           & 410           & 0.630             \\
2.0         & 10         & 41.6           & 729           & 0.514             \\
2.5         & 4          & 4.7           & 77           & 0.541             \\
3.0         & 106        & 5.7            & 117           & 0.457             \\
3.5         & 3          & 5.1            & 82           & 0.349             \\
4.0         & 146        & 3.0            & 66            & 0.422             \\
4.5         & 38         & 2.2            & 52            & 0.462             \\
5.0         & 447        & 2.1            & 48            & 0.437    \\
\hline
\end{tabular}%
        \vspace{5pt}\caption{BFT targets by luminosity class (LC), with median radii (R), distances (d), and angular diameters ($\theta$)}\label{tab-general_targets-B}

    \end{minipage} 
%\caption{BFT general imaging targets.}\label{tab-general_targets}
\end{table}

\section{Architecture and Performance}\label{section_bft-science_performance}

%\subsection{General architecture}\label{section_bft-general_architecture}

These known targets are collectively all brighter than 7.6 in the visible $I$-band, which forms the basis of our performance requirement, with a goal of 8.0 mag.  In the near-infrared $H$-band, the matching requirement and goal are 5.6 and 6.0, respectively.  
These two bands are selected provide a matching pair: one for greatest signal-to-noise, to phase the array with pair-wise aperture combination for real-time fringe tracking (FTK); and one for maximum resolution, to provide the most pixels on target with all-in-one combination, taking advantage of a coherence time synthetically extended through the NIR FTK.  The near-infrared channel is most attractive for sensitivity due to the slower atmospheric coherence time and higher Strehl ratio.

{\bf Beam relay.} To eliminate a major source of both capital construction costs as well as substantial ongoing operations \& maintenance costs, the BFT will employ a vacuum-free design for beam relay from the outboard stations to the beam combination laboratory.  Instead, single-mode polarization-maintaining fibers (SMPM) will be used for the relay task.  Within this context, the two astronomical bands were chosen to match available, low-attenuation SMPM fibers.  For wavelengths shorter than $I$-band, and longer than $H$-band, fiber attenuation increases sharply from the values of $\sim$1-4 db/km (Table \ref{tab-BFT_performance}) for commercially available fiber.  Our current experience base with the VISION instrument \cite{Garcia2016PASP..128e5004G} with Thorlabs PM630-HP SMPM fiber indicates that the related PM780-HP and PM1550-XP items would serve reasonably in this regard.

\begin{figure}
    \centering
    \includegraphics[width=0.95\linewidth]{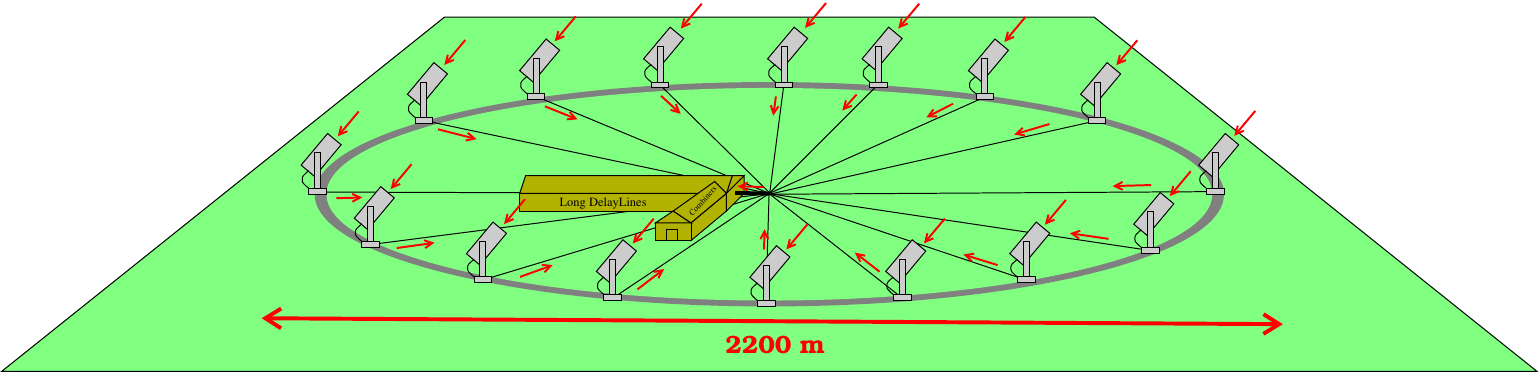}
    \caption{Nominal schematic of the BFT: 16 $\times$ 0.5 meter telescopes on a 2.2-km ring, with a central beam delay and combination laboratory.}
    \label{figure_bft}
\end{figure}

{\bf Overall site size.}  In all of the primary science cases of \S \ref{section_bft-science_case}, target counts noted above are for angular sizes bracketed to be 0.40 to 0.60 milliarcseconds in size, with a desire for snapshot imaging for $10 \times 10$ to $30 \times 30$ pixels. Cursory image reconstruction experiments with the APSYNSIM interferometric imaging simulator \cite{Marti-Vidal2017arXiv170600936M} indicates this is satisfied by an optical telescope with a sparse aperture of 16 elements on a 2,200 m ring.  This level of `snapshot' (no sky rotation needed) imaging is adequate for starspot capture and tracking in latitude and longitude over a stellar rotation and magnetic evolution cycle (days and years, respectively).  The ring count and circumference mean a telescope-to-telescope spacing of $\sim$390 meters, designed for H-band pairwise fringe tracking with sufficient signal-to-noise from adequate fringe visibility for targets up to 0.75mas in size.  The redundancy of the ring geometry means that individual station-to-station pair dropouts do not corrupt the overall baseline bootstrapping of the ring.  `True' imaging (at the Airy criterion of 1.22 $\lambda/D$) will be at 90 $\mu$as pixel resolution, with model-based reconstruction (the Michelson criterion of 0.25 $\lambda/D$) \cite{Kervella2008A&A...488..667K,Martinache2010ApJ...724..464M} at 19 $\mu$as pixel resolution.

{\bf Siderostat geometry.}  Each of the 16 input stations will employ a 0.5 m equivalent aperture, in the form of a siderostat-fed fixed telescope.  A siderostat configuration offers a number of significant advantages for the BFT.  
\begin{itemize}
    \item Articulation of a single large optic (the siderostat flat) rather than an entire telescope tube means the enclosure can be significantly smaller.
    \item With a smaller enclosure, the enclosure design can be a rollaway roof, rather than a more traditional dome/slit arrangement.  The latter arrangement has greater cost and mechanical complexity, making it more prone to failure.  Rollaways also tend to have more forgiving failure modes, with it being more straightforward to manually close a malfunctioning roof.
    \item A fixed telescope means any fiber-feeding optics -- and the attached fiber -- can also remain fixed. Keeping beam relay fibers static improves their performance, and makes characterization of that performance more persistent.
    \item A siderostat configuration allows the facility to be used in full-aperture retroreflection mode, which means internal fringes are possible.  Such a measurement can be used during the day (or night) to monitor facility health and establish internal path length constant terms.
\end{itemize}

There a couple of disadvantages of such a configuration, which can be either mitigated or accommodated.  First, a siderostat configuration means two large optics, rather than one, must be manufactured and maintained (eg. cleaning and coating).  (Interestingly, the downward-pointing nature of the fixed telescope as it faces the siderostat means it is less prone to accumulate dust on the primary mirror surface.)   Second, a siderostat can impose limitations in sky coverage.  This can be mitigate through a slight oversize of the siderostat flat relative to the fixed telescope size - eg. a 0.5 m telescope would require a 0.65 m flat - and this oversize can be relaxed if some vignetting and/or sky coverage reduction is allowed for certain pointings. Facilities such as IOTA \cite{Carleton1994SPIE.2200..152C,Pedretti2008SPIE.7013E..2VP} and PTI \cite{Colavita1999ApJ...510..505C} have employed this architecture for similar reasons.

% \usepackage{graphicx}
\begin{wraptable}{r}{9cm}
\centering
\resizebox{8.8cm}{!}{%
\begin{tabular}{|l|l|l|l|}
\hline
\textbf{Parameter} &
  \textbf{\begin{tabular}[c]{@{}l@{}}H-band\\ (tracking)\end{tabular}} &
  \textbf{\begin{tabular}[c]{@{}l@{}}R-band\\ (science)\end{tabular}} &
  \textbf{units} \\ \hline
Wavelength         & 1.55  & 0.79  & um    \\ \hline
Aperture size      & 0.50  & 0.50  & m     \\ \hline
Strehl ratio       & 0.68  & 0.27  &       \\ \hline
Number of splits   & 2     & 15    &       \\ \hline
Integration time   & 0.009 & 3.0   & s     \\ \hline
Read noise         & 20    & 1     & e$^-$ rms \\ \hline
Optics temperature & 290   & 290   & K     \\ \hline
Object $V^2$          & 0.65  & 0.005 &       \\ \hline
Fiber attenuation  & 1     & 4     & dB/km \\ \hline
Reflections        & 30    & 30    &       \\ \hline
Reflectivity       & 0.975 & 0.975 &       \\ \hline
Optics emissivity  & 0.53  & 0.53  &       \\ \hline
\begin{tabular}[c]{@{}l@{}}Throughput to\\ detector\end{tabular} &
  0.16 &
  0.03 &
   \\ \hline
Detector QE        & 0.70  & 0.70  &       \\ \hline
Sensitivity limit      & 5.95  & 7.89  & mag   \\ \hline
SNR                & 24.30 & 3.00  &       \\ \hline
\end{tabular}%
}
\caption{Engineering parameters for the BFT.}
\label{tab-BFT_performance}
\end{wraptable}

\textbf{Collecting aperture size.}  The overall collecting aperture size of 0.5 m is shown in the next section to be sufficiently large for the science cases noted above.  We additionally find this size is a `sweet spot', for a number of reasons.  First, this size of telescope is readily available from commercial manufacturers, in a cost-effective manner.  Using the metric of `dollars per unit aperture', the 0.5 to 0.6 m diameter apertures are the largest sizes where that metric holds at roughly \$30 per cm$^2$; for 0.7 and 1.0m apertures, this metric starts to increase in a substantial way (which has a daunting aggregate impact if one  is considering purchasing 16 telescopes).  Secondly, at this size, simple tip-tilt target tracking is sufficient for most atmospheric conditions.  Not having to implement higher-order adaptive optics has a substantial impact in increasing reliability, reducing infrastructure cost, simplifying operations, and reducing ongoing maintenance costs.  

\textbf{Beam delay.}  Beam delay represents probably one of the greatest challenges of the BFT, yet also one of the greatest opportunities for taking advantage of new technology.  As noted above, delay lines for CHARA and VLTI are based on technology that dates back to the 80s and 90s \cite{Colavita1992ESOC...39.1143C,Messenger1998Msngr..91...25.}; the more modern units for MROI are still $\sim$15 years old \cite{Fisher2010SPIE.7734E..49F}.  In the interregnum, new options have arisen for active control of beam path alignment and distance to and from moving optical surfaces.  A good example of fine control at levels sufficient for optical interferometry on irregular surfaces is the \textit{Pyxis} ground-based demonstrator for formation-flying optical
interferometry \cite{Hansen2022SPIE12183E..1BH,Hansen2023arXiv230707211H}.  
Our strawman baseline design is a `woofer-tweeter' approach in free-space optics with static long delay lines and dynamic short delay lines, sufficient for a sky coverage from the zenith down to 56$^o$ (to see down the north celestial pole from our notional site) and an hour-plus of continuous tracking.  These delay lines would each employ a multi-pass design, with its multiple reflections accounted for in our sensitivity budget, which would keep costs of a lab enclosure to a reasonable level.  This design element of the BFT is one we are giving the greatest scrutiny, and is highly likely to evolve as we optimize the architecture for performance and cost in the context of currently available technology.   The performance details of our delay line needs are covered below in \S \ref{section_bft-costs}.

\textbf{Beam combination.}  Current instrumentation designs for the BFT are simple notional evolutions of the existing NPOI NIR fringe tracker \cite{Armstrong2018SPIE10701E..0BA} and VISION imager \cite{Garcia2016PASP..128e5004G}.  These concepts could be (and most likely will be) replaced with optimized designs, but for the present represent known, scalable instruments that provide an existence proof that back-end instrumentation for BFT requires no new inventions.  The use of NPOI instrumentation in baseline bootstrapping experiments \cite{Jorgensen2016SPIE.9907E..2CJ} is particularly attractive for BFT implementation.

\section{Derived performance}\label{section_bft-derived_performance}

Typical seeing from a large, flat northern Arizona site is 1.1", corresponding to a $r_0$ for this latter value of 11.4 and 26.6 cm at I- and H-band, respectively \cite{Walters1991BAAS...23..895W,Harris1992PASP..104..140H,Hutter1997AJ....114.2822H}\footnote{Although it is surprisingly difficult to find $\sim$2km$^2$ flat ($<$20m variation), high ($>$1,500m), dark sites ($>$21.7 mag/arcsec$^2$) worldwide, a number of such sites have been identified in northern Arizona and even site tested in previous studies of notional U.S. Cherenkov Telescope Array locations \cite{Ong2013ICRC...33.2848O}. (`Flat' is not required by this kind of interferometric observing but is a significant simplifier for construction and maintenance.)}. For a half-meter telescope, tip-tilt operation with peak tracking should then result in a Strehl of 0.27 and 0.68 in the I- and H-bands, respectively \cite{Glindemann2011psi..book.....G}.  For the H-band fringe tracking, nearest neighbor pairwise combination means each beam is being split 50/50 to combine with those neighbors.  A 5ms coherence time at 500nm is taken as reasonable \cite{Kellerer2007A&A...461..775K}, which scales by $\lambda^{6/5}$ to 8.6ms at 790nm.  Other system parameters: a low-cost InGaAs near-infrared detector with 70\% QE and 20 e$^-$ read noise is baselined (these run $\sim$\$50k, although  ones with significantly better read noise exist, for roughly 10$\times$ the cost); system temperature is room temperature, at 290K, object for short-baseline tracking is $V^2=0.65$ (typical of the fainter objects on a 385m baseline); fiber coupling efficiency is 65\%; fiber attenuation is 1 dB/km; 30 reflections at 0.975 (principally for multi-pass delay lines, and assuming a standard mirror coating such as Thorlabs M01 protected gold).  The reflective losses, fiber coupling \& attenuation, and Strehl multiply together to indicate a throughput to the detector of 16\%.  For high signal-to-noise fringe tracking at a SNR of 24.3, our limiting magnitude for H-band fringe tracking is expected to be $m_H = 5.95$.  A similar computation can be done for the I-band science imaging, with some adjustments.  The fiber attenuation is greater, 4 dB/km; the object will be fully resolved with $V^2$ will be significantly lower, at $\sim$0.005; the Strehl will be lower at the shorter wavelength (noted above); and the number of splits for 16-way combination is far greater.  However, with the H-band fringe tracking, we note a SNR of 3.0 (which scales with wavelength, and the square root of the number of apertures being bootstrapped), yet we then expect a synthetic coherence time of 3 seconds to be achieved, allowing a limiting magnitude for the science band to be $m_I = 7.9$.

These numbers are consistent with our science demands laid out in \S \ref{section_bft-science_performance}.  It is also important to note that these performance evaluations are \textit{empirical} in nature, having been checked against actual on-sky performance numbers of facilities such as CHARA, NPOI, and PTI.

%\needspace{6em}
\section{Projected BFT Costs}\label{section_bft-costs}

The BFT has been under consideration by our group as representative of a next-generation capability that is attainable without an order of magnitude increase in cost -- rather, a step back in the opposite direction is required by budget realities.  This concept is useful in that it captures the current state-of-the-art for the various subsystems, and motivates the need for improvements with transport and delay.  A summary roll-up of these costs can be seen in Table \ref{tab-BFT_cost}.

\textit{Beam Collection.}  Published, off-the-shelf costs for 20-inch PlaneWave observatory systems (optical tube and mount) are \$56k; for each station we are also bookkeeping an additional \$50k for commercially provided siderostat flat and mounting modification for a siderostat configuration.  A 11'$\times$11' rollaway enclosure is bookkept at a known, off-the-shelf cost of \$20k, installed.  Fast tip-tilt is budgeted at \$50k; foundation work at \$15k; telescope pier at \$2.5k; fiber injection optics at \$50k; pointing model detectors at \$2.5k; computing infrastructure is \$2.5k; wireless internet relays are \$2.5k; and off-grid power systems at \$30k.  All of these costs are from direct experience building single-aperture facilities in use at Lowell (the robotic 0.5 m TiMo, robotic 1.0 m PJ1M, the robotic 4$\times$14" JPL NEO survey telescope).  Overall each telescope station will run \$281k per unit.  By focusing on the bright star case, substantial capital cost savings is realized: individual off-the-shelf 1.0m stations would be \$700k or more for the telescope alone, and require full AO systems (at least \$300k each for sufficient performance).

% \usepackage{graphicx}
\begin{wraptable}{r}{7.0cm}
\centering
\resizebox{6.9cm}{!}{%
\begin{tabular}{lr}
\textbf{Element}              & \textbf{Cost (\$k)} \\
Beam Collection               & \$4,720       \\
Beam Transport                & \$2,744       \\
Beam Combination              & \$4,140       \\
Beam Delay                    & \$4,000       \\
Infrastructure                & \$1,943       \\
Labor                         & \$5,250       \\
Total (with 25\% contingency) & \$28,496     
\end{tabular}
}
\vspace{3pt}\caption{\small Construction cost summary for the BFT.}
\label{tab-BFT_cost}
\end{wraptable}

\textit{Beam Transport.}  A substantial cost element for optical interferometers that operate into the visible, such as CHARA and NPOI, is the use of vacuum pipes for beam relay.  These costs include both capital construction as well as operations \& maintenance.  Our baseline approach is to completely eliminate this expensive infrastructure.  Instead, the BFT will utilize SMPM fibers for beam relay.  Each station-to-lab link will utilize a 3-component optical link: a SMPM fiber for fringe tracking at $H$-band, another SMPM fiber for science imaging in $I$-band, and a third for pathlength metrology.  Fibers are roughly \$26 per meter each, with the pipe being roughly \$32 per meter installed in a trench.  Each beam also has a pathlength-monitoring metrology, which is estimated at \$50k per beam.  Overall each station's beamline, at 1,100m in length, is expected to have a capital cost of roughly \$172,000 (and have substantially lower O\&M costs than vacuum).

\textit{Beam Delay.}  
The beam delay has three straightforward, yet demanding requirements, which are (a) range, and (b) rate, and (c) accuracy.  For a facility with a $\sim2.2$km gross size,  sky coverage to a minimum zenith angle of 56$^o$ (the minimum required to get to the north celestial pole from norther Arizona latitudes) requires a accessible delay range of 1,750m.  With six passes into and out of a long delay line of length 150m, a range of 1,800m is accomplished.  A building able to accommodate a space of length 150m is sizeable but not unreasonable; the CHARA delay line building is 100m in length.  The east-west projection of that 2.2km baseline -- what the delay line must track -- is changing most rapidly as targets transit the meridian; this results in a maximum tracking rate of 153mm/sec.  With six passes into and out of a continuous delay line of physical length 45m, a tracking time of 60 minutes is achievable in the worst case.  For accuracy, such a delay line would need to have an accuracy of 5nm during tracking, resulting in an aggregate pathlength control accuracy of 60nm, meeting an optical quality spec of 1/10 of a wave at 600nm.  Our cost target for each station's delay line is \$250,000 per unit; we have begun risk-reduction lab efforts that are intended to demonstrate the achievability of this cost target.  Previous generation delay line installations have had similar per-line costs.

\textit{Beam Combination.}
As a notional baseline for our point design, beam combination will employ two proven designs.  For the near-infrared fringe tracker (NIRFTK), we will duplicate the NPOI pairwise beam combiner 
\cite{Armstrong2018SPIE10701E..0BA}, scaled up from 6 to 16 beams; for a pairwise combiner, such scaling is linear.  This design will be further improved with a lower read noise detector, graduating from a CRED2 to a CRED1.  Similarly, the NPOI VISION combiner \cite{Garcia2016PASP..128e5004G} can be scaled up, although this scales as $N^2$, rather than linearly.  As with \textit{Beam Collection}, above, this represents a known solution and, for the purposes of this paper, we will not be focusing on improvements.  Overall, the expected capital cost of this backend, based on known numbers from the NPOI NIRFTK and VISION combiners, will be \$4,140k.

\textit{Infrastructure.}  The cost of delay enclosures, an instrument wing, and control wing is straightfoward; in particular, the delay enclosure will be based on commercially available warehouses, but largely empty due to our new delay line design.  Site access and site power is also included in this cost element, which totals just under \$2,000k.

\textit{Labor.}  For the facility construction, 10 FTEs per year for 3 years is baselined, including 6 techs, 2 engineers, and 2 scientists, resulting in an expected fully burdened labor cost of \$5,250k.  Importantly, by using a significant number of COTS components in conjunction with the elements demonstrated by this proposal, a well-defined \& limited build schedule can be realistically established for a next-generation optical interferometer.

\textit{Operations \& Maintenance.}  Beyond capital construction, an utterly essential component of any next-generation interferometry facility is a high degree of automation, robustness, and health monitoring, that result in reduced O\&M costs.  The baseline BFT architecture calls for these features to be designed into each subsystem from the start, as well as fully robotic facility operations, as was achieved at the Mark III interferometer, and partially achieved at the Palomar Testbed Interferometer facility.  Our proposed developments for \textit{Beam Transport} and  \textit{Beam Delay} above will be aimed not only at hitting the necessary performance targets, but economic ones as well.
Within that context, an annual operating budget of \$750k is targeted for a BFT telescope, which includes a staff of 2 full-time techs, and half a FTE for both an engineer and an operations scientist.  By focusing on such costs from the project inception, this represents a $>4\times$ reduction in O\&M costs relative to current (and smaller) optical interferometers.  The intention is to have a explicitly defined operational life, with a sunset date of 15 years after its commencement of operations, meaning that full life cycle costs can be forecast from the inception of the project.

Overall, realistic costs and schedule are intended to be assessed and controlled throughout the BFT project. Having designs assessed for not just up-front capital and installation costs, but explicit assessments of operations and maintenance costs, will be essential to conducting the full lifecycle of the facility within an achievable budget.  Historically, out-year operations budgets for interferometric observatories have been fueled by crippling underestimates and wishful thinking.  The intention is for BFT to be well-characterized and planned in this regard before proceeding from design to implementation.

\section{Summary}

Bright star imaging is an area ripe for discovery.  The  mature technology for such discovery is commercially available, translating to rapid implementation at reasonable, well-defined cost.  A focus of scientific capability that is carefully balanced against affordable recurring operations \& maintenance costs mean that the BFT will be able to carry out that noteworthy program of discovery on a defined program cost and schedule.  The project is currently at the stage of risk-reduction exercises (eg. on beam delay \& transport), and well as pre-Phase A planning and detailed costing.

\bibliographystyle{spiebib2b}
\bibliography{journal-references}

\end{document}